\documentclass[
								reprint,
								aps,
								prl,
								nofootinbib,
								twocolumn,
								notitlepage,
							]{revtex4-1}

\usepackage{lmodern}
\usepackage{amsmath,amssymb,amsthm,amsfonts,mathrsfs}
\usepackage{graphicx,epstopdf,epsfig,enumerate}
\usepackage{color}
\usepackage[colorlinks=true,linkcolor=blue,citecolor=blue,urlcolor=blue]{hyperref}

\DeclareGraphicsExtensions{.eps}

\begin{document}

\title{Comments on the paper ``Foundation of statistical mechanics from symmetries of entanglement" by S. Deffner and W. \.Zurek}

\author{Robert Alicki}
\email[e-mail: ]{fizra@ug.edu.pl}
\affiliation{Institute of Theoretical Physics and Astrophysics, University of Gda\'nsk, 80-952 Gda\'nsk, Poland}

\begin{abstract}
The authors of the recent paper \cite{Zurek} boldly claim to discover a new fully quantum approach to foundation of statistical mechanics:\\ 
\emph{ Our conceptually novel approach is free of mathematically ambiguous notions such as probability, ensemble, randomness, etc. }.\\
 The aim of this note is to show that this approach is neither specific for quantum systems nor really conceptually different from the standard textbook arguments supporting microcanonical or canonical ensembles in statistical mechanics.
\end{abstract}

\maketitle
Despite the long history the foundations of statistical mechanics are still a subject of debate, often of the rather philosophical nature. Recently, a large number of papers devoted to this topic appeared, mostly promoting the ideas of quantum information in application to statistical physics. The recent paper of Deffner and \.Zurek \cite{Zurek} contains particularly  bold claims promising  elimination from foundations of quantum statistical mechanics `` mathematically ambiguous notions such as probability, ensemble, randomness, etc.". Such controversial statement deserves a deeper debate and the aim of this short note is to point out certain issues  which make questionable the arguments of \cite{Zurek}.\\

\textbf{1) From microcanonical to canonical equilibrium state}\\

The presented derivation of the canonical equilibrium state for a small subsystem of a large total system being in the microcanonical equilibrium state is  standard and can be found
for example in the textbook \cite{Lopuch} pp. 101-105. It is enough to replace classical discrete cells by quantum energy eigenstates. Therefore, only the authors arguments concerning the origin of microcanonical state are discussed here.\\

\textbf{2) Definition of microcanonical state}\\

The authors of \cite{Zurek} confront their definition:\\

\emph{A) The microcanonical equilibrium of a system S with Hamiltonian $H_S$ is an even
(envariant under all unitaries), fully energetically degenerate quantum state;}\\

with the ``standard" one :\\

\emph{B) A macroscopic system samples every permissible quantum state with equal probability.}\\

First of all the authors definition makes sense only for the trivial Hamiltonian $H_S =0$ for which the ``standard" definition can be rephrased as:\\

\emph{C) The microcanonical equilibrium of a system S with a Hamiltonian $H_S =0$  is the state invariant under all unitaries.}\\

The only difference between the definitions A) and C) is that A) uses a purification of a state while C) a mixed state of a system. The ``envariance" of purification is equivalent to the invariance of the original state. The  use of pure quantum states does not eliminate the notion of probability which is intrinsic for  quantum mechanics. It is also unclear why the definition A) should be recognized as natural from physical or mathematical point of view.

The philosophical questions related to  the notion of probability (e.g. "ensembles versus single system interpretation") are the same for classical and quantum theory and the  notions of purification and entanglement are not helpful in this matter. On the other hand it is very controversial to call the notion of probability ``mathematically ambiguous". 
Since the Kolmogorov axiomatization   ``probability theory is measure theory with a soul" ( Mark Kac, see also \cite{Kac} for interesting views on this topic).\\

\textbf{3) Degeneracy of Hamiltonian spectra}\\

By this opportunity one should mention that the postulate of massive degeneracy of  Hamiltonian spectra for large quantum systems used in \cite{Zurek} and recently also in many papers on `` resource theory" approach to statistical mechanics is highly unphysical. Generic Hamiltonians have essentially non-degenerated spectra, any massive degeneracy means the presence of additional constants of motion beside energy which spoil  ergodicity of a system. On the other hand ergodicity is a necessary condition for thermodynamical behavior. Therefore, the really standard quantum microcanonical state is a normalized projection on a subspace spanned by  Hamiltonian eigenvectors with energies in a  certain ``small" but macroscopic interval. In fact, a better definition involves a projection on a subspace spanned by  Hamiltonian eigenvectors with energies smaller than a certain fixed value. The later state is passive \cite{Woronowicz} and in the thermodynamical limit is, anyway, equivalent to the former one.\\

\textbf{4) Classical meaning of ``envariance"}\\

In their approach the authors of \cite{Zurek} use "entanglement assisted invariance - envariance" which is supposed to be an entirely quantum property. It is correct only if we restrict ourselves to classical pure states (points in a phase-space). On the other hand quantum pure states have many properties of classical mixed states like non-zero transition probability between different quantum pure states. A better analog of a quantum pure state, which can be always treated  as an eigenvector of a certain ``Hamiltonian" $H$ to  an eigenvalue $E$, is the  microcanonical probability distribution traditionally expressed as Dirac delta  $\delta( H(x) - E)$. Here, $H(x)$ is the classical Hamiltonian corresponding to the quantum one and $x$ denotes  phase-space variables. The convergence of  energy eigenstates to classical microcanonical distributions has been proved for a number of quantum Hamiltonians (see e.g. \cite{Helffer}) what supports this picture of  quantum pure states. Introducing another  classical system described by  variables $y$ and a Hamiltonian $H'(y)$ one can consider a joint microcanonical state $\delta [(H(x) + H'(y)) - E]$.  One can say that this state is ``envariant" if there exist two canonical transformations $g,g'$ such that  $\delta [(H(g(x)) + H'(y)) - E] = \delta [(H(x) + H'(g'(y))) - E]$. In particular if the first Hamiltonian is trivial, $H(x) \equiv 0$ then averaging over $y$ one obtains a uniform probability distribution over $x$ - a ``microcanonical state".\\

The author thanks Adam Majewski for discussions.

\end{document}